\documentclass[aps,tightenlines,prl,a4paper,twocolumn,superscriptaddress,showpacs,amsmath,amssymb,preprintnumbers,footinbib,floatfix]{revtex4}

\usepackage{graphicx}
\usepackage{dcolumn}

\usepackage{array}
\usepackage{color}

\def\bz{B^0}
\def\bzkx{\bz \to K^{*0} X} 
\def\bzkxmu{B^0 \to K^{*0} X,~K^{*0}\to K^+ \pi^-,~X \to \mu^+ \mu^-}
\def\kxmumu{B^{0}_{K^{*}X}}
\def\bzrhox{\bz \to \rho^{0} X}
\def\bzrhoxmu{\bz \to \rho^{0} X,~\rho^{0} \to \pi^+ \pi^-,~X \to \mu^+ \mu^-}
\def\rhoxmumu{B^{0}_{\rho X }}
\newcommand{\mbc}{{M_{\textrm{bc}}}}

\newcommand{\BR}{{\mathcal B}}

\newcommand{\gev}{{\hbox{GeV}}}
\newcommand{\pgev}{{\hbox{GeV}/c}}
\newcommand{\mmev}{{\hbox{MeV}/c^2}}
\newcommand{\mgev}{{\hbox{GeV}/c^2}}

\begin{document}
  
    \preprint{\vbox{
         \hbox{KEK Preprint 2010-6}
        \hbox{BELLE Preprint 2010-8}
       }
    }
  
\title{Search for a Low Mass Particle Decaying into $\mu^+ \mu^-$ in \boldmath{$\bzkx$} and \boldmath{$\bzrhox$} at Belle}

\affiliation{Budker Institute of Nuclear Physics, Novosibirsk}
\affiliation{Faculty of Mathematics and Physics, Charles University, Prague}
\affiliation{University of Cincinnati, Cincinnati, Ohio 45221}
\affiliation{Department of Physics, Fu Jen Catholic University, Taipei}
\affiliation{Justus-Liebig-Universit\"at Gie\ss{}en, Gie\ss{}en}
\affiliation{Hanyang University, Seoul}
\affiliation{University of Hawaii, Honolulu, Hawaii 96822}
\affiliation{High Energy Accelerator Research Organization (KEK), Tsukuba}
\affiliation{Institute of High Energy Physics, Chinese Academy of Sciences, Beijing}
\affiliation{Institute of High Energy Physics, Protvino}
\affiliation{Institute for Theoretical and Experimental Physics, Moscow}
\affiliation{J. Stefan Institute, Ljubljana}
\affiliation{Kanagawa University, Yokohama}
\affiliation{Korea Institute of Science and Technology Information, Daejeon}
\affiliation{Korea University, Seoul}
\affiliation{Kyungpook National University, Taegu}
\affiliation{\'Ecole Polytechnique F\'ed\'erale de Lausanne (EPFL), Lausanne}
\affiliation{Faculty of Mathematics and Physics, University of Ljubljana, Ljubljana}
\affiliation{University of Maribor, Maribor}
\affiliation{Max-Planck-Institut f\"ur Physik, M\"unchen}
\affiliation{University of Melbourne, School of Physics, Victoria 3010}
\affiliation{Nagoya University, Nagoya}
\affiliation{Nara Women's University, Nara}
\affiliation{National Central University, Chung-li}
\affiliation{National United University, Miao Li}
\affiliation{Department of Physics, National Taiwan University, Taipei}
\affiliation{H. Niewodniczanski Institute of Nuclear Physics, Krakow}
\affiliation{Nippon Dental University, Niigata}
\affiliation{Niigata University, Niigata}
\affiliation{University of Nova Gorica, Nova Gorica}
\affiliation{Novosibirsk State University, Novosibirsk}
\affiliation{Osaka City University, Osaka}
\affiliation{Panjab University, Chandigarh}
\affiliation{Saga University, Saga}
\affiliation{University of Science and Technology of China, Hefei}
\affiliation{Seoul National University, Seoul}
\affiliation{Sungkyunkwan University, Suwon}
\affiliation{School of Physics, University of Sydney, NSW 2006}
\affiliation{Tata Institute of Fundamental Research, Mumbai}
\affiliation{Excellence Cluster Universe, Technische Universit\"at M\"unchen, Garching}
\affiliation{Tohoku Gakuin University, Tagajo}
\affiliation{Tohoku University, Sendai}
\affiliation{Department of Physics, University of Tokyo, Tokyo}
\affiliation{Tokyo Metropolitan University, Tokyo}
\affiliation{Tokyo University of Agriculture and Technology, Tokyo}
\affiliation{IPNAS, Virginia Polytechnic Institute and State University, Blacksburg, Virginia 24061}
\affiliation{Yonsei University, Seoul}
  \author{H.~J.~Hyun}\affiliation{Kyungpook National University, Taegu} 
  \author{H.~K.~Park}
	\email[Corresponding author. Email: ]{hkpark@knu.ac.kr}
	\affiliation{Kyungpook National University, Taegu} 
  \author{H.~O.~Kim}\affiliation{Kyungpook National University, Taegu} 
  \author{H.~Park}\affiliation{Kyungpook National University, Taegu} 
  \author{H.~Aihara}\affiliation{Department of Physics, University of Tokyo, Tokyo} 
  \author{K.~Arinstein}\affiliation{Budker Institute of Nuclear Physics, Novosibirsk}\affiliation{Novosibirsk State University, Novosibirsk} 
  \author{T.~Aushev}\affiliation{\'Ecole Polytechnique F\'ed\'erale de Lausanne (EPFL), Lausanne}\affiliation{Institute for Theoretical and Experimental Physics, Moscow} 
  \author{A.~M.~Bakich}\affiliation{School of Physics, University of Sydney, NSW 2006} 
  \author{E.~Barberio}\affiliation{University of Melbourne, School of Physics, Victoria 3010} 
  \author{A.~Bay}\affiliation{\'Ecole Polytechnique F\'ed\'erale de Lausanne (EPFL), Lausanne} 
% ...
\collaboration{The Belle Collaboration}
  
\begin{abstract}
We search for dimuon decays of a low mass particle in the decays $\bz \to K^{*0} X $ and $\bz \to \rho^0 X$ using a data sample of
$657 \times 10^6~B \bar B$ events collected with the Belle detector at the KEKB asymmetric-energy $e^+ e^-$ collider.  
We find no evidence for such a particle in the mass range from 212 $\mmev$ to 300 $\mmev$, and set
upper limits on its branching fractions. In particular, we search for a particle
with a mass of 214.3 $\mmev$ reported by the HyperCP experiment, and obtain upper limits on the products 
$\BR(\bzkx)\times\BR(X \to \mu^+ \mu^-)< 2.26~(2.27) \times 10^{-8}$ and $\BR(\bzrhox)\times\BR(X \to \mu^+ \mu^-)< 1.73~(1.73) \times 10^{-8}$ at 90\% C.L. 
for a scalar (vector) $X$ particle.
\end{abstract}
  
  
  \maketitle
      {
	\renewcommand{\thefootnote}{\fnsymbol{footnote}}}
      \setcounter{footnote}{0}

The possibility of a weakly interacting light particle with a mass from a few MeV to a few GeV has been extensively discussed~\cite{theomodel}. Recent astrophysical observations by PAMELA~\cite{pamela} 
and ATIC~\cite{atic} have been interpreted as dark matter annihilation
mediated by a light gauge boson, called the $U$-boson~\cite{hamed}, which couples to Standard Model particles. 
In addition, the HyperCP collaboration~\cite{hkpark} has reported three $\Sigma^+ \to p \mu^+ \mu^-$ events with 
dimuon invariant masses clustered around 214.3 $\mmev$ that are consistent with the process $\Sigma^+ \to p X, X \to \mu^+ \mu^-$.
Phenomenologically, $X$ could either be a pseudoscalar or an axial-vector particle~\cite{he1} with a lifetime for the 
pseudoscalar case estimated to be about $10^{-14}$ s~\cite{geng}.
Many plausible explanations for such a particle have been proposed; a pseudoscalar sgoldstino particle~\cite{sgoldstino} 
in various supersymmetric models~\cite{susy}, a light pseudoscalar Higgs boson~\cite{higgs} 
in the Next-to-Minimal-Supersymmetric Standard Model as well as a vector $U$-boson~\cite{uboson} as described above. 

Recently there have been searches for a similar light particle at the Tevatron~\cite{D0}, $e^+ e^-$ colliders~\cite{cleo} 
and fixed-target experiments~\cite{E391, KTeV}. In those searches, the light particle was assumed to be a pseudoscalar and 
no evidence has been found. The KTeV result in $K_L$ decay disfavors a pseudoscalar explanation of the HyperCP results~\cite{KTeV}. 

The large sample of $B^0$ decays at the Belle provides a good opportunity to search for
a light scalar or vector particle.
In particular, the estimated branching fractions for $B^0 \to V X, X \to \mu^+ \mu^-$
where $X$ is a sgoldstino particle with a mass of 214.3 $\mmev$ and $V$ is either a $K^{*0}$ or $\rho^0$ meson, 
are in the range $10^{-9}$ to $10^{-6}$~\cite{demidov}.

We report a search for a light particle using the modes, $\bzkxmu$ ($\kxmumu$) and $\bzrhoxmu$ ($\rhoxmumu$) 
using a data sample of $657 \times 10^6~B \bar B$ pairs collected with the Belle detector~\cite{belled} 
at the KEKB asymmetric-energy $e^+ e^-$ collider~\cite{KEKB}.
The analysis for $\kxmumu$ uses the same dataset as Ref.~\cite{JTWei_PRL}. 
In this analysis, we assume that the light $X$ particle is either a scalar or vector particle. 
Unless specified otherwise, charge-conjugate modes are implied. The term scalar (vector) $X$ particle implies either a 
scalar (vector) or pseudoscalar (axial-vector) particle throughout this letter.

The Belle detector is a large-solid-angle magnetic spectrometer that consists of a silicon vertex detector (SVD),
a 50-layer central drift chamber (CDC), an array of aerogel threshold Cherenkov counters (ACC),
a barrel-like arrangement of time-of-flight scintillation counters (TOF), and an electromagnetic calorimeter (ECL)
comprised of CsI(Tl) crystals located inside a superconducting solenoid coil that provides a 1.5~T
magnetic field. An iron flux-return located outside of the coil is instrumented to detect $K_L^0$ mesons and to identify 
muons (KLM). 

In the initial event selection, at least two oppositely charged muon tracks with momenta larger than 0.690 $\pgev$
are required. These muon tracks are selected using a likelihood ratio formed from a combination of the track penetration depth 
and hit pattern in the KLM system.
We reduce the number of badly reconstructed tracks by requiring that $|dz|<5.0$ cm and $dr < 1.0$ cm, where $|dz|$ and $dr$
are distances of closest approach of a track to the interaction point in the beam direction ($z$) and 
in the transverse plane ($r-\phi$), respectively. 
Charged kaons and pions are identified using information from the ACC and TOF systems and the energy loss ($dE/dx$) measurements 
in the CDC~\cite{BellePID}.
  
The reconstruction of $K^{*0}$ ($\rho^0$) in the $\kxmumu$ ($\rhoxmumu$) decay 
uses identified $K^+$ ($\pi^+$) and  $\pi^-$ ($\pi^-$) tracks.
The reconstructed invariant mass $M_{K^{*0}}$ ($M_{\rho^0}$) of $K^{*0}$ ($\rho^0$) candidates for the decay mode 
$\kxmumu$ ($\rhoxmumu$) is required to be in the ranges 0.815 $\mgev$ $<$ $M_{K^{*0}}$ $<$ 0.975 $\mgev$ 
(0.633 $\mgev$ $<$ $M_{\rho^0}$ $<$ 0.908 $\mgev$), corresponding to $\pm1.5 \sigma$ ($\pm1 \sigma$) in the reconstructed mass distribution. 
The $\mu^+ \mu^-$ dimuon tracks are used to reconstruct low mass $X$ candidates.  

$\kxmumu$ ($\rhoxmumu$) candidates are reconstructed from a  $K^{*0}$ ($\rho^0$) candidate and a pair of muons. 
Reconstructed $\bz$ candidates are selected using the beam-energy-constrained mass
$\mbc = \sqrt{E_\textrm{beam}^2-p_{B}^2}$ and energy difference $\Delta E = E_B - E_\textrm{beam}$, 
where $E_\textrm{beam}$ is the beam energy and $E_B$ ($p_B$) are the energy (momentum) of the reconstructed $\bz$ candidates 
evaluated in the center-of-mass frame. $\bz$ candidates are required to lie in the signal regions, 
5.27 $\mgev$ $<$ $\mbc$ $<$ 5.29 $\mgev$ and $-0.03$ $\gev$ $<$ $\Delta E$ $<$ 0.04 $\gev$ ($-0.04$ $\gev$ $<$ $\Delta E$ $<$ 0.04 $\gev$) 
for the decay $\kxmumu$ ($\rhoxmumu$).
In events containing more than one $\bz$ candidate, we select the best $\bz$ candidate with the smallest $\chi^2$ value, 
where $\chi^2$ is obtained when the four charged tracks are fitted to a common vertex.
Using this algorithm, we select the correct $\kxmumu$ and $\rhoxmumu$ combinations in the $\mbc$ and $\Delta E$ signal region 96.6\% (96.7\%) and 93.7\% (93.5\%) of the time for a scalar (vector) $X$ particle, respectively. 
The signature for $X \to \mu^+ \mu^-$ in $\kxmumu$ and $\rhoxmumu$ decays 
would be a peak in the dimuon mass.
The width of the signal region for the light particle search with mass below 300 $\mmev$ is $3\sigma$ in dimuon mass resolution.
The dimuon mass resolutions for $\kxmumu$ and $\rhoxmumu$ vary from 0.5 $\mmev$ to 1.9 $\mmev$
as the mass of $X$ ($M_X$) increases from 212 $\mmev$ to 300 $\mmev$.
However, the signal region for dimuon mass ($M_{\mu \mu}$) for 214.3 $\mmev$ of the HyperCP event search 
is defined to be 211.6 $\mmev$ $<$ $M_{\mu \mu}$ $<$ 217.2 $\mmev$ where the width of the search region is $\pm 3 \sigma$ 
in the combined mass resolution, which is obtained by  
linearly summing the mass resolutions of the HyperCP and Belle detectors.

For background studies, we employ two different techniques referred to as the counting ($\mathcal{C}$) and fitting ($\mathcal{F}$) methods. 
Method $\mathcal{C}$ uses generic $B \bar{B}$ and continuum ($e^+ e^- \to q \bar q,~q=u, d, s, c$)
Monte Carlo (MC) samples that correspond to an integrated luminosity about three times larger than the data sample. 
In the $\Delta E-\mbc$ signal region, there are no events in the dimuon mass region $M_{\mu \mu}$ $<$ 225 $\mmev$
($M_{\mu \mu}$ $<$ 239 $\mmev$) for the decay $\kxmumu$ ($\rhoxmumu$). 
In method $\mathcal{F}$, we use the MC samples as described above, and select $\bz$ candidates in the
sideband regions defined as $-0.12$ $\gev$ $<$ $\Delta E$ $<$ $-0.06$ $\gev$ and 0.06 $\gev$ $<$ $\Delta E$ $<$ 0.12 $\gev$, and 
5.25 $\mgev$ $<$ $\mbc$ $<$ 5.27 $\mgev$. By fitting 
the dimuon mass distributions for the $\bz$ candidates with a probability density function, 
$(x-0.21)^n$ for $x > 2 m_{\mu}$, where $x$ is a dimuon mass in $\mgev$, $m_{\mu}$ is the muon mass and the parameter $n$ is extracted from the fit, 
we estimate the number of background events
with dimuon mass below 300 $\mmev$. We also compare the shape of the probability density function with the $\bz$ candidates 
in data sideband regions. No significant discrepancy is found. 
The estimated numbers of background events for methods $\mathcal{C}$ and $\mathcal{F}$
for the HyperCP event search are $0$ ($0$) and $0.13^{+0.04}_{-0.03}$ ($0.12^{+0.03}_{-0.02}$)
for the decays $\kxmumu$ ($\rhoxmumu$), respectively.
The background estimates for both methods give results that are equivalent within statistical errors for masses below 300 $\mmev$.

\begin{figure}[htbp]
    \includegraphics[height=43mm,width=81mm]{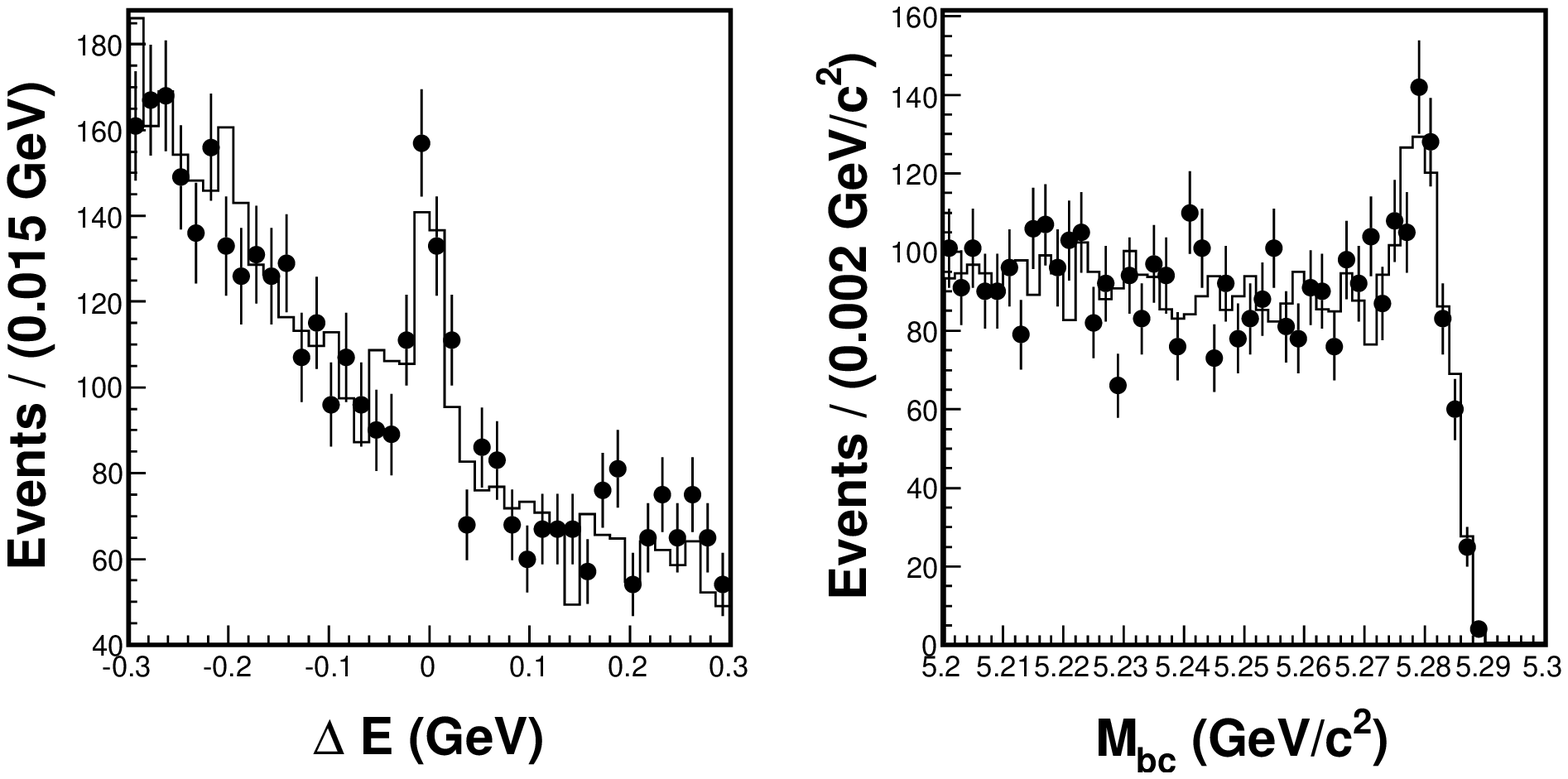}
    \includegraphics[height=43mm,width=81mm]{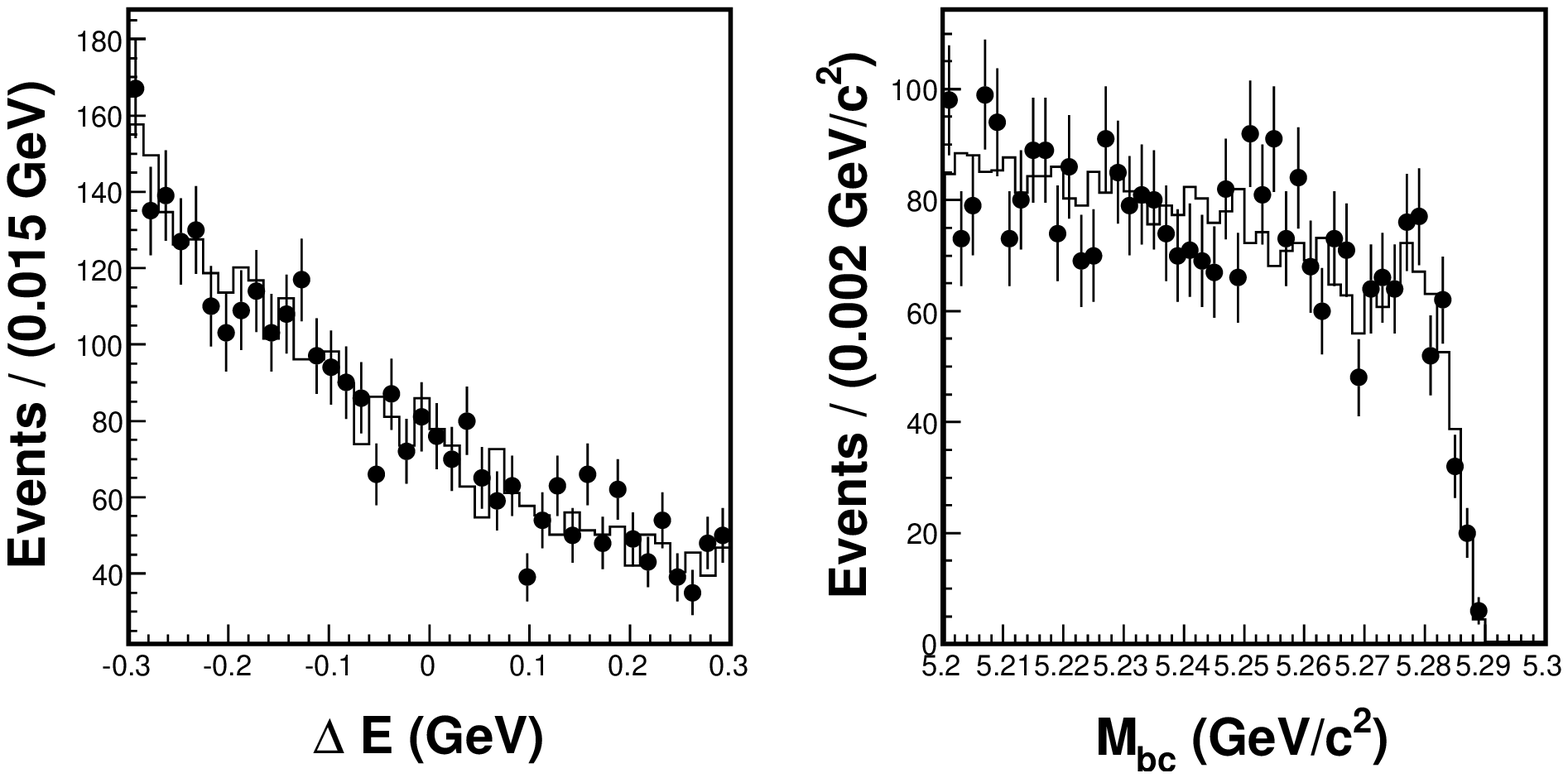}
    \caption{Data and MC comparison for $\Delta E$ and $\mbc$ distributions for 
	$\kxmumu$ (top) and $\rhoxmumu$ (bottom) candidates. The points with error bars and histograms represent data and background MC, respectively.}
   \label{fig:damccomp}
\end{figure}

\begin{figure}[htbp]
    \includegraphics[height=42mm,width=65mm]{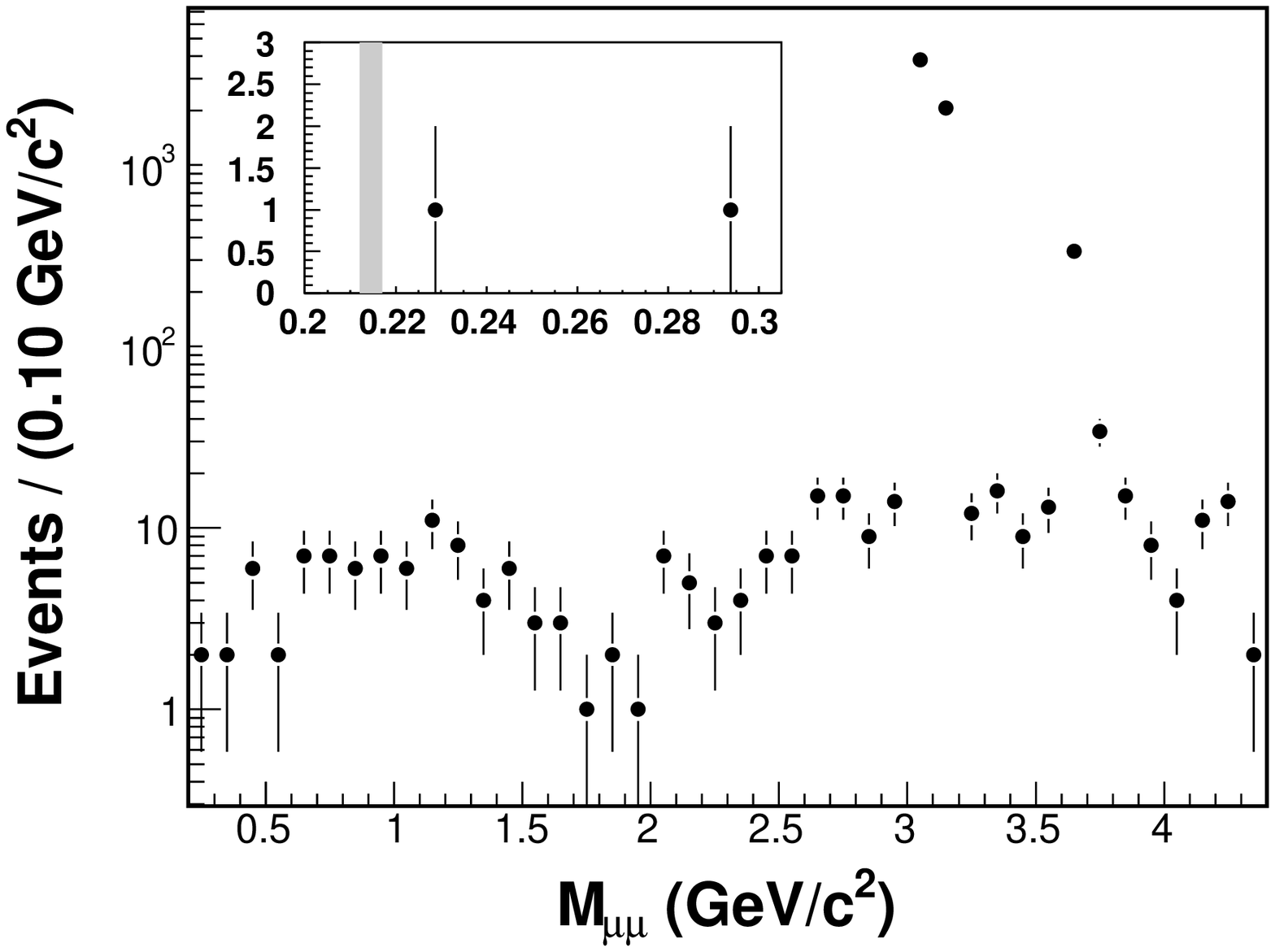}
    \includegraphics[height=42mm,width=65mm]{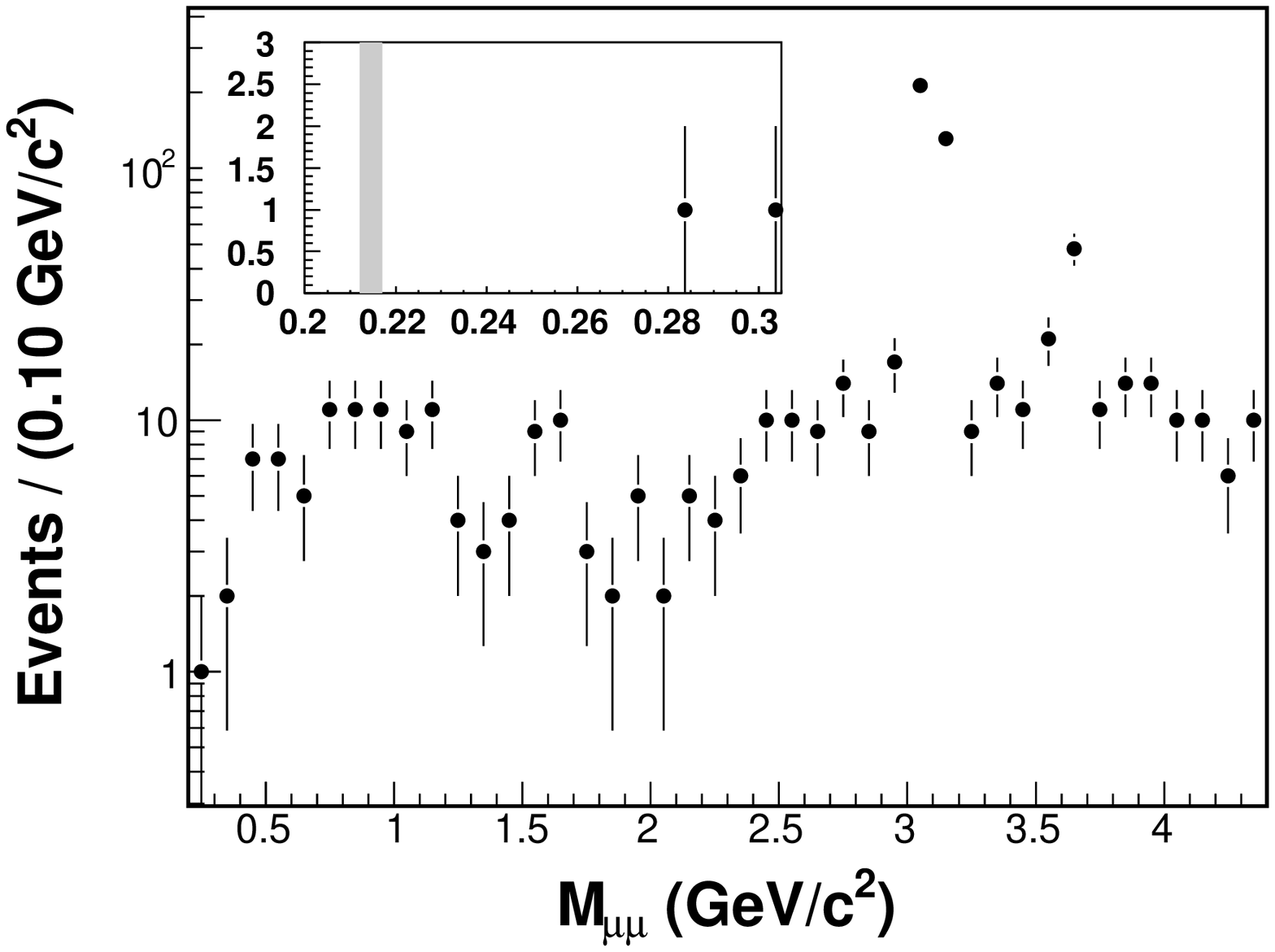}
    \caption{Dimuon mass distribution for the $\kxmumu$ (top) and $\rhoxmumu$ (bottom) candidates in the
    signal regions for $\mbc$ and $\Delta E$. The shaded region in the inset shows the HyperCP mass region.}
   \label{fig:dimuon}
\end{figure}

Before examining the full data sample, various distributions, including $\mbc$, $\Delta E$, 
dimuon mass and $dz$ in the background MC samples are compared with a small fraction of the data.
These are in good agreement. Figure~\ref{fig:damccomp} shows the data and MC comparison for
$\Delta E$ and $\mbc$ distributions after the best $\bz$ candidates are selected. 
The peaks in the $\Delta E$ and $\mbc$ distributions for the $\kxmumu$ are mainly due to $B^0 \to J/\psi K^{*0},~J/\psi \to \mu^+\mu^-$.
The dimuon mass distributions including the $J/\psi$ and $\psi'$ mass regions for $\kxmumu$ and $\rhoxmumu$ candidates 
in the signal regions of $\mbc$ and $\Delta E$ are shown in Fig.~\ref{fig:dimuon}.
There are no events observed in the HyperCP mass region.

For the full data sample, no significant signal is observed for the decays $\kxmumu$ and $\rhoxmumu$ for $M_{X}$ below
$\sim 300$ $\mmev$. We derive an upper limit for the signal yield ($S_{90}$) at a 90\% confidence level (C.L.)  
by using the POLE program~\cite{POLE} with the Feldman-Cousins method~\cite{FCmethod}. 
This procedure takes into account Poisson fluctuations 
in the number of observed signal events and Gaussian fluctuations in the estimated number of background events 
as well as systematic uncertainties.
The $S_{90}$ values for the HyperCP event search are $2.33$ ($2.33$) for $\kxmumu$ decay with a scalar (vector) $X$ 
and $2.33$ ($2.33$) for $\rhoxmumu$ decay with a scalar (vector) $X$ particle.
 
\begin{table*}
\begin{centering}
\caption{Summary of the number of observed events ($N_{obs}$), estimated number of background events ($N_{bg}$), efficiencies ($\epsilon$), signal yields ($S_{90}$) 
and upper limits ($U.L.$) at 90\% C.L. 
for the decays $\kxmumu$ and $\rhoxmumu$ with the scalar (vector) $X$ particle. The errors on $N_{bg}$ are statistical only.}
\label{tab:summ}
\begin{footnotesize}
\begin{ruledtabular}
\renewcommand{\arraystretch}{1.3}
\begin{tabular}{c|ccccc|ccccc}
~~~$M_{\mu\mu}$ & \multicolumn{5}{c|}{$\bzkxmu$} & \multicolumn{5}{c}{$\bzrhoxmu$}  \\ 
   ($\mmev$)    & $N_{obs}$ & $N_{bg}$ & $\epsilon$ & $S_{90}$ & $U.L. (10^{-8})$   
                                                                     & $N_{obs}$ & $N_{bg}$ & $\epsilon$ &$S_{90}$ & $U.L. (10^{-8})$ \\ 
\hline
212.0           & 0 & $0.03^{+0.01}_{-0.01}$ ($0.03^{+0.01}_{-0.01}$) & 23.8 (23.7)& 2.43 (2.43)& 2.34 (2.34)     

		& 0 & $0.02^{+0.01}_{-0.01}$ ($0.02^{+0.01}_{-0.01}$) & 21.2 (21.1)& 2.44 (2.44)& 1.77 (1.78) \\

214.3           & 0 & $0.13^{+0.04}_{-0.03}$ ($0.13^{+0.04}_{-0.03}$) & 23.6 (23.5)& 2.33 (2.33)& 2.26 (2.27)     

		& 0 & $0.12^{+0.03}_{-0.02}$ ($0.12^{+0.03}_{-0.02}$) & 20.7 (20.7)& 2.33 (2.33)& 1.73 (1.73) \\ 

220.0           & 0 & $0.13^{+0.02}_{-0.02}$ ($0.13^{+0.02}_{-0.02}$) & 23.0 (22.9)& 2.33 (2.33)& 2.31 (2.33)     

		& 0 & $0.11^{+0.02}_{-0.01}$ ($0.11^{+0.02}_{-0.01}$) & 20.2 (20.1)& 2.33 (2.33)& 1.78 (1.78) \\ 

230.0           & 1 & $0.24^{+0.02}_{-0.02}$ ($0.25^{+0.02}_{-0.02}$) & 21.4 (21.4)& 4.09 (4.12)& 4.37 (4.40)     

		& 0 & $0.21^{+0.01}_{-0.01}$ ($0.21^{+0.01}_{-0.01}$) & 18.8 (18.9)& 2.27 (2.27)& 1.86 (1.85) \\

240.0           & 0 & $0.38^{+0.02}_{-0.02}$ ($0.39^{+0.02}_{-0.02}$) & 20.0 (20.0)& 2.09 (2.09)& 2.40 (2.39)     

		& 0 & $0.32^{+0.01}_{-0.01}$ ($0.32^{+0.01}_{-0.01}$) & 17.5 (17.5)& 2.16 (2.16)& 1.90 (1.90) \\                                                                             
250.0           & 0 & $0.51^{+0.01}_{-0.01}$ ($0.51^{+0.01}_{-0.01}$) & 18.0 (18.4)& 1.92 (1.94)& 2.43 (2.41)     

		& 0 & $0.42^{+0.00}_{-0.00}$ ($0.42^{+0.00}_{-0.00}$) & 15.9 (16.3)& 2.06 (2.06)& 1.99 (1.94) \\                                                                                                             
260.0           & 0 & $0.63^{+0.01}_{-0.01}$ ($0.63^{+0.01}_{-0.01}$) & 16.5 (17.2)& 1.83 (1.83)& 2.54 (2.43)     

		& 0 & $0.60^{+0.01}_{-0.00}$ ($0.70^{+0.01}_{-0.00}$) & 14.5 (15.2)& 1.84 (1.80)& 1.95 (1.82) \\   

270.0           & 0 & $0.75^{+0.02}_{-0.02}$ ($0.75^{+0.02}_{-0.02}$) & 15.4 (16.4)& 1.76 (1.76)& 2.61 (2.45)     

		& 0 & $0.61^{+0.02}_{-0.01}$ ($0.61^{+0.02}_{-0.01}$) & 13.7 (14.4)& 1.83 (1.83)& 2.06 (1.96) \\

280.0           & 0 & $0.69^{+0.03}_{-0.03}$ ($0.86^{+0.04}_{-0.04}$) & 14.6 (15.8)& 1.78 (1.69)& 2.78 (2.45)     

		& 1 & $0.83^{+0.03}_{-0.03}$ ($0.90^{+0.04}_{-0.03}$) & 13.0 (13.9)& 3.52 (3.45)& 4.17 (3.83) \\

290.0           & 1 & $0.98^{+0.06}_{-0.06}$ ($0.97^{+0.06}_{-0.06}$) & 14.0 (15.5)& 3.35 (3.37)& 5.47 (4.99)     

		& 0 & $0.80^{+0.04}_{-0.04}$ ($0.78^{+0.04}_{-0.04}$) & 12.4 (13.6)& 1.74 (1.74)& 2.16 (1.97) \\

300.0           & 1 & $1.08^{+0.08}_{-0.08}$ ($1.08^{+0.08}_{-0.08}$) & 13.6 (15.1)& 3.28 (3.28)& 5.53 (4.97)     

		& 1 & $0.87^{+0.05}_{-0.05}$ ($0.87^{+0.05}_{-0.05}$) & 11.9 (13.3)& 3.48 (3.48)& 4.51 (4.01) \\
\end{tabular}
\end{ruledtabular}
\end{footnotesize}
\end{centering}
\end{table*}

Upper limits on the branching fraction for the decays $\kxmumu$ and $\rhoxmumu$ are obtained from  

\begin{eqnarray}
\BR (\bz \to V X, X \to \mu^+ \mu^-)< \frac{S_{90}}{\epsilon \times N_{B \bar B} \times \mathcal{B}_{V}}, \nonumber
\label{eq:br}
\end{eqnarray}

\hspace{-0.15in}where $V$ stands for either $K^{*0}$ or $\rho^0$, and $\mathcal{B}_V$~\cite{pdg} are the intermediate vector meson branching fractions, 
$\BR(K^{*0} \to K^+ \pi^-$) or $\BR(\rho^0 \to \pi^+ \pi^-)$.
Here $N_{B \bar B}$ and $\epsilon$ denote the number of $B \bar B$ pairs and the signal efficiency with small data/MC corrections for 
charged particle identification, respectively.

The signal efficiency is determined by applying the same selection criteria to the signal MC sample as those used for the data.
The signal MC samples for a scalar (vector) $X$ particle
are generated for $X$ masses in the range 212 $\mmev$ $\leq$ $M_X$ $\leq$ 300 $\mmev$
using the $P \to V S$ ($P \to V V$) model in the EvtGen generator~\cite{evtgen} for a scalar (vector) $X$ particle. 
In the MC generation of the vector $X$ particle, we assume that the polarization of $X$ is either
fully longitudinal or transverse. The efficiency differences between longitudinal and transverse polarizations of the $X$ 
for both modes in the search range are less than 7 \%. 
Since the efficiencies for a fully longitudinal polarized $X$ are lower than for a fully transversly polarized $X$,
we conservatively use the efficiencies for full longitudinal polarization of the $X$ for upper limit estimations.      
In the HyperCP event search for a scalar (vector) $X$ particle, the efficiencies 
for $\kxmumu$ and $\rhoxmumu$ decays are 23.6\% (23.5\%) and 20.7\% (20.7\%), respectively.
We also check the efficiencies for different $X$ lifetimes.  
The efficiencies are the same for lifetimes below $10^{-12}$ s because the primary and secondary vertices are indistinguishable.
The efficiencies for the two different vertex fitting methods for the HyperCP event search are compared.  
One method assumes that the dimuon tracks from the $X$ originate from the primary $\bz$ decay vertex, while the other assumes   
that the dimuon tracks from the $X$ are from a secondary vertex. The difference in the efficiencies is about 1 \%.

\begin{table}[htpb]
\caption{Summary of fractional systematic uncertainties in the upper limit for a scalar (vector) $X$ particle in the HyperCP mass range 
for the decays $\kxmumu$ and $\rhoxmumu$, respectively.}
\label{tab:syst}
\begin{ruledtabular}
\renewcommand{\arraystretch}{1.1}
\begin{tabular}{lrr}
  & \multicolumn{2}{c} {$\sigma_{\BR}/\BR$ (\%) \rule[-1.5mm]{0mm}{5mm}} \\
\multicolumn{1}{c}{\raisebox{1.5ex}[0pt]{Source}}  
  & \multicolumn{1}{c}{$\kxmumu$}     
  & \multicolumn{1}{c}{$\rhoxmumu$} \\ \hline
$N_{B \bar B}$                         &  1.4 (1.4) &  1.4 (1.4)  \\
$\mu^{\pm}$ identification             &  4.2 (4.2) &  4.1 (4.1)  \\
$K^\pm$ identification                 &  0.8 (0.8) &  -          \\
$\pi^\pm$ identification               &  0.5 (0.5) &  1.0 (1.0)  \\
Tracking efficiency                    &  4.2 (4.2) &  4.4 (4.3)  \\
$\mbc$                                 &  0.5 (0.3) &  0.3 (0.6)  \\
$\Delta E$                             &  0.5 (0.3) &  0.3 (0.6)  \\
$K^{*0}$ tagging                       &  0.5 (0.3) &  -          \\
$\rho^0$ tagging                       &  -         &  0.3 (0.6)  \\
MC statistics			       &  0.1 (0.1) &  0.1 (0.1)  \\\hline
Total                                  &  6.2 (6.2) &  6.2 (6.3)  \\ 
\end{tabular}
\end{ruledtabular}
\end{table}

To obtain the final upper limit, we use the backgrounds determined from the fitting method.
Since the efficiencies for a scalar (vector) and a pseudoscalar (axial-vector) are the same, 
the upper limits for the scalar (vector) and the pseudoscalar (axial-vector) $X$ searches are identical. 
From the $\kxmumu$ ($\rhoxmumu$) sample, the upper limits for a scalar and vector $X$ particle in the HyperCP mass range are determined to be
$2.26~(1.73) \times 10^{-8}$ and  $2.27~(1.73) \times 10^{-8}$, respectively.
Table~\ref{tab:summ} summarizes the number of observed events, the expected number of background events, the efficiencies, the signal yields,  
and the upper limits at 90\% C.L. in the interval 212 $\mmev$ $\leq$ $M_X$ $\leq$ 300 $\mmev$. 

The systematic uncertainties in the upper limits for the decays $\kxmumu$ and $\rhoxmumu$ 
in the HyperCP mass range are summarized in Table~\ref{tab:syst}. 
The total systematic uncertainties in the upper limits for both decay modes
vary from 6\% to 8\% as the mass of $X$ increases from 212 $\mmev$ to 300 $\mmev$. 
The dominant systematic uncertainties come from tracking efficiency and muon identification. 
The uncertainty for the tracking efficiency is estimated by linearly summing the single track systematic errors, which are $\sim 1 \%$/track. 
The uncertainty of muon identification is measured as a function of momentum and direction 
by using the $\gamma \gamma \to \mu^+ \mu^-$ data sample.

In summary, we searched for a scalar and vector particle in the decays
$\bzkxmu$ and $\bzrhoxmu$ in the mass region 212 $\mmev$ $\leq$ $M_X$ $\leq$ 300 $\mmev$. 
No significant signals are observed in a sample of $657 \times 10^6~B \bar B$ pairs. 
We set 90\% C.L. upper limits of $\BR(\bzkxmu) < 2.26 \times 10^{-8}$ $(2.27 \times 10^{-8})$
and $\BR(\bzrhoxmu) < 1.73 \times 10^{-8}$ $(1.73 \times 10^{-8})$ for a 214.3 $\mmev$ mass scalar (vector)
$X$ particle; our results rule out models II and III for the sgoldstino interpretation of the HyperCP observation~\cite{demidov}.

\begin{acknowledgments}
We thank the KEKB group for excellent operation of the
accelerator, the KEK cryogenics group for efficient solenoid
operations, and the KEK computer group and
the NII for valuable computing and SINET3 network support.
We acknowledge support from MEXT, JSPS and Nagoya's TLPRC (Japan);
ARC and DIISR (Australia); NSFC (China); MSMT (Czechia);
DST (India); MEST, NRF, NSDC of KISTI and WCU (Korea); MNiSW (Poland);
MES and RFAAE (Russia); ARRS (Slovenia); SNSF (Switzerland);
NSC and MOE (Taiwan); and DOE (USA). 
H. Park acknowledges support by NRF Grant No. R01-2008-000-10477-0.
\end{acknowledgments}

\end{document}